\documentclass{article}

\usepackage[english]{babel}

\usepackage[letterpaper,top=2cm,bottom=2cm,left=3cm,right=3cm,marginparwidth=1.75cm]{geometry}

\usepackage{amsmath}
\usepackage{graphicx}
\usepackage[colorlinks=true, allcolors=blue]{hyperref}
\newtheorem{lemma}{Lemma}
\newtheorem{theorem}{Theorem}

\title{Automatic Increase Market Systems (AIMS): Towards a deterministic theory for cryptocurrencies}
\author{ Wantall Newby, Nickuk Nishikawa}

\begin{document}
\maketitle

\begin{abstract}
The popularity of cryptocurrencies has grown significantly in recent years, and they have become an important asset for internet trading. One of the main drawbacks of cryptocurrencies is the high volatility and fluctuation in value. The value of cryptocurrencies can change rapidly and dramatically, making them a risky investment. Cryptocurrencies are largely unregulated, which can exacerbate their volatility. The high volatility of cryptocurrencies has also led to a speculative bubble, with many investors buying and selling cryptocurrencies based on short-term price fluctuations rather than their underlying values. Therefore, how to reduce the fluctuation risk introduced by exchanges, transform uncertain prices to deterministic value, and promote the benefits of decentralized finance are critical for the future development of cryptos and Web 3.0. 

To address the issues, this paper proposes a novel theory as Automatic Increase Market Systems (AIMS) for cryptos, which could potentially be designed to automatically adjust the value of a cryptocurrency helping to stabilize the price and increase its value over time in a deterministic manner. We build a crypto, \textsf{WISH} (https://wishbank.wtf),  based on AIMS in order to demonstrate how the automatic increase market system would work in practice, and how it would influence the supply of the cryptocurrency in response to market demand and finally make itself to be a stable medium of exchange, ensuring that the AIMS is fair and transparent. \newline

\end{abstract}

\section{Introduction}
Cryptocurrencies are virtual tokens that use cryptography for security and operate independently of a central bank~\cite{mukhopadhyay2016brief}. They are decentralized and use blockchain technology to keep track of transactions, making them transparent and immutable~\cite{nakamoto2008bitcoin}.
Many businesses and individuals now accept cryptocurrencies as a form of payment, and there are numerous exchanges and trading platforms that allow users to buy, sell, and trade cryptocurrencies~\cite{fang2022cryptocurrency}. However, it is important to note that cryptocurrencies are still a relatively new and volatile asset, and there are huge risks involved in investing or trading cyrptos~\cite{trautman2022ftx,gupta2022empirical,nagpal2017cryptocurrency}. The high volatility of cryptos may lead to significant losses for investors who are not prepared for price fluctuations. And the fluctuations make cryptos less attractive for use as a medium of exchange. Many merchants might be hesitant to accept cryptocurrencies due to the high volatility, which can result in significant losses if the value falls sharply. The result would distinctly discourage the adoption of the corresponding decentralized applications~\cite{gupta2022empirical,bellon2022bubbles}. 

It is difficult to completely eliminate the speculative nature of cryptocurrencies and the associated risks, as market forces and human behavior are inherently unpredictable, especially over unregulated exchanges~\cite{jalan2023systemic} and crypto projects~\cite{jacobs2018cryptocurrencies,hairudin2022cryptocurrencies}. However, there have been proposals for various systems and mechanisms to reduce the volatility of cryptocurrencies and address some of the risks associated with them.
For example, 
an automatic market maker (AMM) system could help to stabilize the price of a cryptocurrency by adjusting the supply and demand of the currency in response to market conditions. An AMM system uses algorithms to set the price of a cryptocurrency based on the amount of liquidity available in the market, and adjusts the supply of the currency in response to changes in demand~\cite{egorov2021automatic}.
Another potential approach is the use of stablecoins, which are cryptocurrencies that are pegged to the value of a fiat currency or other asset~\cite{fiedler2023stablecoins}. 
In addition, increased regulation and oversight of the cryptocurrency market could help to reduce the risks associated with cryptocurrency investment and trading. This could include measures such as requiring exchanges to comply with anti-money laundering and know-your-customer regulations, and establishing clearer guidelines for the issuance and trading of cryptocurrencies~\cite{corbet2019cryptocurrencies}.
It is important to note that the volatility of the current design of cryptocurrencies and the associated risks are inherent to the nature of the market, which determines that the above solutions have no guarantee to eliminate the risks~\cite{jalan2023systemic}. 

This paper proposes a novel theory as Automatic Increase Market Systems (AIMS) for cryptocurrencies, which could potentially be designed to automatically adjust the value of a cryptocurrency. The adjustment is contracted on blockchains to be transparent, which means that the increase of its value over time is deterministic regardless of the exchanges and market conditions. Thus, 
the AIMS is a blockchain-based mechanism that aims to create a new type of cyrpto financial market where the value of a cryptocurrency only increases over time. This is achieved by using an automated system deployed with a smart contract that adjusts the price of the cryptocurrency in response to a deterministic mathematical function, ensuring that the price of the cryptocurrency only goes up.
The use of AIMS could help to reduce of the risks, stabilize the prices and make cryptocurrencies a more deterministic, stable and reliable asset for investment and trading. The contribution of the paper is threefold as follows:
\begin{itemize}
\item To propose and formulate the Automatic Increase Market Systems (AIMS) for cryptocurrencies which enables  deterministic value change of a cryptocurrency and complete decentralized exchange.   
\item To summarize the features and benefits towards using Automatic Increase Market Systems (AIMS) with analysis comparing with traditional cryptocurrencies.
\item To demonstrate the use, design, and implementation of Automatic Increase Market Systems (AIMS) for \textsf{WISH} (https://wishbank.wtf) in practice. Source code of the smart contract and deployed application are both available for public  reference.
\end{itemize}

The rest of the paper is organized as follows: Section~\ref{problem} presents the concepts and theories of AIMS, and Section~\ref{demo}
describe the \textsf{WISH} crypto project implementation. Section~\ref{related} reviews the literature, and  Section~\ref{conclude} finally
 concludes the paper.
 
\begin{figure}
\centering
\includegraphics[width=0.6\textwidth]{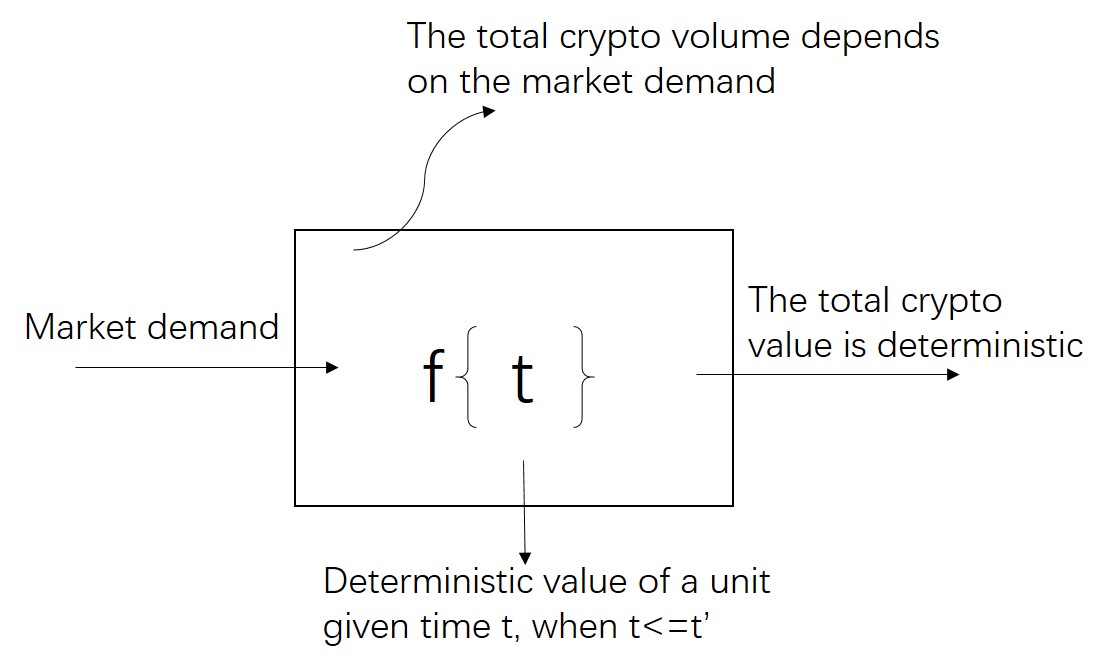}
\caption{\label{fig:problem}The illustration of an AIMS contract.}
\end{figure}

\section{Automatic Increase Market Systems (AIMS)}
\label{problem}
Let $P$ be the set of all participants in a blockchain network $N$, where each participant $i$ has a public key $\pi$ and a private key $\psi$. Let $T$ be the set of all transactions in the network. Let $C$ be the smart contract code deployed on $N$, which is a program that executes automatically when certain conditions are met. Let $\gamma$ be the USDT-similar  stablecoin on $N$. 

A cryptocurrency $c$ issued by $C$ is denoted as $\Delta$. The value of $\Delta$ is a deterministic increase function $f_{C}()$ of time $t$, where $t$ is a variable within a timespan ($t\in(t_i,t')$) as the input to the function, namely we have the formula as follows,
\[
\Delta_{C,t} = f_{C}(t), t\in(t_i,t'),
\]
where $t_i$ is an initial time point to automatically produce the price of $c$ regards to $\gamma$, and $t'$ is the ultimate time point for the use of $f_{C}()$. We can gain insights into its properties and characteristics from Figure~\ref{fig:problem}, which shows that the total crypto volume and value are determinstic to the market demand as an input of $C$ for $f_{C}()$ within a timespan.  

Given such a function $f_{C}()$ for $c$ on $N$,  we have the following lemmas.
\begin{lemma}
Let $t_x<t_y$, $t_x,t_y\in{t_i,t'}$, we have $\Delta_{C,t_x}<\Delta_{C,t_y}$. 
\end{lemma}
The lemma shows the deterministic increase feature of $c$ with AIMS. The $c$ value on $N$ always increases with time within the timespan according to a given function.  
\begin{lemma}
Given the amount of $\gamma$ to be $\xi$ for a participant $i\in P$ on $N$, the following formula always hold,
\[
\frac{\xi}{\Delta_{C,t}} \geq 0.
\]
\end{lemma}
The lemma holds for a non-negative value of $\xi$, and the produced price of $c$ at $t$ shall be always positive. And the coined cryptos over the time $T$ for a participant $i$ is 
$\sum_{t\in T}{\frac{\xi}{\Delta_{C,t}}}$.
\begin{lemma}
Given the amount of $\gamma$ to be $\xi_t$ for a participant $i\in P$ at time $t$ on $N$, the total locked value $\Omega>0$ of $C$ on $N$ for $i$ at time $t_m$ is,
\[
	\Omega = \sum_{t\in T}{\frac{\xi_t}{\Delta_{C,t}}} \times f_{C}(t_m).
\]
\end{lemma}
The total locked value of $c$ shows the potential profit-earning ability of a participant $i$. Thus for all the participants the total locked value is $\sum_{t\in T}^{P}{\frac{\xi_t}{\Delta_{C,t}}}$, which is the total market value of $C$. When $t>t'$, the increase slope of $\Omega$ is 0. 

\begin{lemma}
Given the amount of $\gamma$ to be $\xi_t$ for a participant $i\in P$ at time $t$ on $N$, the total net profits $\Lambda$ of $C$ on $N$ within $T$ ($t<t_m>T$) for $P$ is,
\[
	\Lambda = \sum_{t\in T}^{P}{\frac{\xi_t}{\Delta_{C,t}}} \times (f_{C}(t_m)-\Delta_{C,t}).
\]
\end{lemma}

\begin{theorem}
Given the amount of $\gamma$ to be $\xi_t$ for a participant $i\in P$ at time $t$ on $N$, we have
\[
\Omega > \Lambda.
\]
\end{theorem}

Let $\Xi $ is a function of activities encouraging the participants on $N$ to destroy $c$, then we have Theorem~\ref{the_last} that we shall always have a $\Xi$ to reach a balance $0$ of the total locked value regards to the cryptocurrency net profits. And it means that it is possible for the a cryptocurrency to reach to a stable status without inflated price according to the invested values in $\gamma$ and to provide a pure function of stabilizing the prices. However, to realize such a $\Xi$, a significant effort has to be invested. 

\begin{theorem}
\label{the_last}
Given the amount of $\gamma$ to be $\xi_t$ for a participant $i\in P$ at time $t$ on $N$, $\Xi$ to be a function of a set activities of destroying $c$,
\[
\Xi = \Omega - \Lambda \neq \emptyset.
\]
\end{theorem}

\section{wishbank.wtf Demonstration}
\label{demo}
In this section, we will show a demonstration of AIMS with a cryptocurrency \textsf{WISH} (https://wishbank.wtf) to illustrate the mechanism in a straightforward way. According to the previous sections, \textsf{WISH} is designed to have several key features as follows. 
\begin{itemize}
\item Automated investment management:  AIMS uses  algorithms to help investors make more informed investment decisions and optimize their trading strategies. This includes features such as automated rebalancing, risk management, and portfolio optimization.
\item Automatic increase market system: The AIMS system uses an automated increase market system to passively adjust the volume of the cryptocurrency in response to market conditions and demand. This ensures that the price and supply of the cryptocurrency are totally transparent, providing a more predictable investment option.
\item Decentralized exchange: AIMS operates as a decentralized exchange, which means that trades are executed directly on the blockchain and users retain control over their own private keys and funds. This helps to ensure the security and transparency of trades, and eliminates the need for intermediaries such as centralized exchanges.
\end{itemize}

\begin{figure}
\centering
\includegraphics[width=0.6\textwidth]{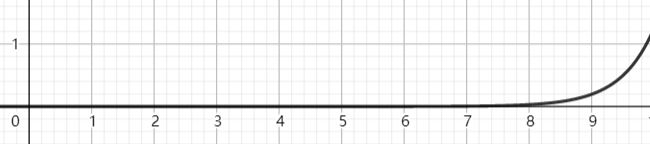}
\caption{\label{fig:plot}The function of value plot with time of \textsf{WISH}}
\end{figure}

For the automation of the increase management of \textsf{WISH},  $f_{C}()$ is defined as 
\[
6.4428653^{n}, n~is~a~unit~of~year~within~[2023~Mar ~6th, ~2033~Jan~21st]
\]
where $n$ is a unit of a year or 365 days from the start of 2023 Mar 6th. Given the initial price of 
\$0.00000001 for a \textsf{WISH}, the price at $n$ is $f_{C}()\times \$0.00000001$. 
Therefore, we can see that for \textsf{WISH} the increase function of its value is a power function of time with base $6.4428653$ multiplying its initial value. The value of $f_{C}()$ depends on the value of time $n$. When $n$ is a positive integer, raising $6.4428653$ to the power of $n$ means multiplying 6.4428653 by itself $n$ times.  The function for the timespan of near 10 years to 2033 Jan 21st as plotted in Figure~\ref{fig:plot}. As you can see from the graph, the function starts at $\$0.00000001$ and increases rapidly as $n$ increases. By the time 2033 Jan 21st, the value of the function is approximately stablizes at \$1.00000005841. The curve is smooth and continuous, indicating that the function is well-behaved over the interval, and the rate of increase slowing down as $n$ gets larger.

The logy growth of the value enables sufficient time for the community to leverage the applications of the cryptocurrency in a relatively long term instead of afraid of the rapid price fluctuation. For \textsf{WISH} few activities $\Xi$ including donation and wish redeem are provided for users to destroy the coins. 

Finally, for the time after 2033 Jan 21st, we can observe that the price and value of \textsf{WISH} is stable to be \$1.00000005841 which can be leveraged as a trust medium of exchange, especially when the term is sufficiently long and the potential growth profit is low.

\section{Literature Review}
\label{related}
Cryptocurrency has gained increasing attention in recent years due to its potential to transform the traditional financial system. A literature review was conducted to explore the current state of research on cryptocurrency, its history, technical aspects, and potential applications. The history of cryptocurrency dates back to the late 1990s when the concept of digital currencies was first introduced. However, it was not until the launch of Bitcoin in 2009~\cite{nakamoto2008bitcoin} that cryptocurrency gained mainstream attention. Since then, numerous cryptocurrencies have been developed, each with its own unique features and use cases~\cite{dutta2022outliers}. 
From a technical perspective, cryptocurrency relies on blockchain technology to create a decentralized and secure system for digital transactions. The blockchain is a distributed ledger that records all transactions and is maintained by a network of computers, making it nearly impossible to hack or manipulate~\cite{radziwill2018blockchain}. This feature has led to the emergence of numerous decentralized applications and smart contracts, which are built on top of blockchain technology. The potential applications of cryptocurrency are vast and varied, ranging from online purchases and international money transfers to voting systems and secure record-keeping. However, the adoption of cryptocurrency has been hindered by several challenges, including regulatory issues, lack of mainstream acceptance, and concerns over security and volatility~\cite{fang2022cryptocurrency}.

In recent years, researchers have focused on addressing these challenges and exploring new use cases for cryptocurrency. Some studies have examined the effectiveness of different consensus algorithms, such as proof-of-stake, in ensuring the security and reliability of blockchain networks. Others have explored the potential for cryptocurrency to improve financial inclusion, particularly in underbanked and developing countries. Despite the potential of cryptocurrency, there are still many questions and uncertainties surrounding its future. Ongoing research and development are necessary to address the challenges and explore new use cases for this transformative technology.

To sum up, cryptocurrency is a rapidly evolving technology with many open questions. Its history, technical aspects, and potential applications have been explored in various studies, but many challenges and uncertainties remain. Further research and development are necessary to fully understand and harness the power of cryptocurrency~\cite{voshmgir2019token,bis2018looking}.

\section{Conclusion and Future Work}
\label{conclude}
AIMS has several benefits for traditional cryptocurrencies, including
a) Reduced volatility. The automatic increase market system used by AIMS  helps to reduce the volatility of the cryptocurrency market by providing a more stable and predictable investment option.
b) Increased transparency. The decentralized exchange used by  AIMS helps to increase transparency and reduce the potential for market manipulation.
c) Enhanced security. The use of blockchain technology and decentralized exchanges helps to enhance the security of trades and ensure the integrity of the system.
Overall, AIMS represents a novel new approach and theory to cryptocurrency design and investment that could have significant benefits for investors and help to address some of the challenges and risks associated with traditional cryptocurrency markets.

In the future, decentralized exchanges and asset swaps including NFTs~\cite{chen2022toward,bamakan2021decentralized} among  blockchains are possible to be purely designed on AIMS, which introduces less risks and higher transparency compared to the existing state of the art for cryptocurrencies.

\bibliographystyle{alpha}
\bibliography{sample}

\end{document}